\documentclass[pre,showpacs,twocolumn]{revtex4}
\usepackage{epsfig}
\usepackage{dcolumn}
\usepackage{bm}

\newcommand{\be}{\begin{equation}}
\newcommand{\ee}{\end{equation}}
\newcommand{\bc}{\begin{center}}
\newcommand{\ec}{\end{center}}
\newcommand{\bi}{\begin{itemize}}
\newcommand{\ei}{\end{itemize}}
\newcommand{\ba}{\begin{eqnarray}}
\newcommand{\ea}{\end{eqnarray}}

\newcommand{\ignore}[1]{}
\newcommand{\mean}[1]{\left\langle #1 \right\rangle}
\newcommand{\abs}[1]{\left| #1 \right|}

\begin{document}
\title{Ising model on two connected Barabasi-Albert networks}
\author{Krzysztof Suchecki}
\email{suchecki@if.pw.edu.pl} \affiliation{Faculty of Physics and
Center of Excellence for Complex Systems Research \\Warsaw
University of Technology \\ Koszykowa 75, PL--00-662 Warsaw,
Poland}
\author{Janusz A. Ho{\l}yst}
\email{jholyst@if.pw.edu.pl} \affiliation{Faculty of Physics and
Center of Excellence for Complex Systems Research \\Warsaw
University of Technology \\ Koszykowa 75, PL--00-662 Warsaw,
Poland}
\date{\today}

\begin{abstract}
We investigate analytically the behavior of Ising model on two connected Barabasi-Albert networks. Depending on
relative ordering of both networks there are two possible phases corresponding to parallel or antiparallel
alingment of spins in both networks. A difference between critical  temperatures of both phases disappears in the
limit of vanishing inter-network coupling for identical networks. The analytic predictions are confirmed by
numerical simulations.
\end{abstract}
\maketitle

\section{Introduction}
Ising model on Barabasi-Albert (B-A) scale-free network \cite{barabasi} has been investigated both numerically
\cite{staufer} as well as analytically \cite{bianconi} and it has been shown that the critical temperature of such
a system is proportional to the logarithm  of a total spins number $N$. Similar Ising models have been
investigated for a general class of random scale-free graphs \cite{Dorog,critical,herrero} and it has
been shown that critical temperatures of such systems depend substantially on a characteristic exponent $\gamma$
describing the probability distribution of node degrees.
Other studies of this model include investigations of antiferromagnetic interactions \cite{antiferro}, dynamics on
directed networks \cite{directed} and critical properties of spin-glass \cite{spinglass}.\\ Besides a large
interest in physical properties of Ising-like models they seem also to be important for opinion formation modeling
\cite{social,social2,galam,nowak}. The Ising model exhibits a majority rule dynamics - the feature that often can be
found in social systems, where a given person changes his/her opinion to fit to a  majority of his neighbors.
Since it is common that social networks have modular structure of weakly coupled clusters \cite{community} it is
of particular interest to see how the Ising model behaves in the case of two interacting complex networks. While
geometrical properties of interconnected complex networks have been studied before \cite{growing}, the dynamics in
such systems has not been explored throughly.\\ We start with  analytical investigations  of Ising model on two
interconnected Barabasi-Albert networks, and then show results of numerical simulations confirming our analytical
studies.

\section{Model}
Our model considers Ising spins on a B-A network. Each node of the network has a spin. We study only ferromagnetic
interactions existing between directly connected spins.

\begin{figure}[ht]
 \vskip 0.5cm
 \centerline{\epsfig{file=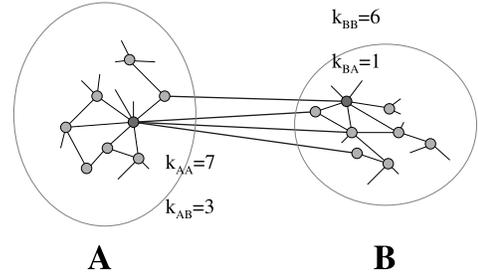,width=.8\columnwidth}}
 \caption{\label{rys_sieci}Two connected B-A networks. A few nodes from each network are shown. The intra-network degrees $k_{AA}$ and $k_{BB}$ as well as inter-network degrees $k_{AB}$ and $k_{BB}$ for two sample nodes are presented.}
\end{figure}

The B-A model is a model of a growing network \cite{barabasi}. One starts with $m$ fully connected nodes, and adds
new nodes to the network. Each new node creates $m$ connections to the extisting network. The probability that a
connection will be made to a node $i$ is proportional to its degree $k_{i}$. This results in a scale-free network,
with a degree distribution $P(k)\sim k^{-3}$.\\ We assume that  two B-A networks are connected by $E_{AB}$ links
(Fig.\ref{rys_sieci}). Each of these links connects a node in network $A$ with a node in network $B$. The chance
to choose a given node as the end of the inter-network link is proportional to the intra-network node degree. This
means that for a small number of links $E_{AB}$, an inter-network degree $k_{AB}$ of a node in the network $A$ is
proportional to a intra-network node degree $k_{AA}$. A similar relation holds for degrees in the network $B$.
\label{dysk}

\section{Analytic calculations}
The problem of the Ising model in a single B-A network was already solved analytically by Bianconi by an
appropriately tailored mean-field approach \cite{bianconi}. We use a similar approach for the problem of two
connected networks.\\ The Hamiltonian of the Ising model for a single B-A network can be written as \be
H=-\sum_{i,j} J_{ij} s_i s_j -\sum_{i} h_i s_i \label{startowe} \ee where $s_i,s_j=\pm 1$ are spins of nodes
$i,j$, a constant $J_{ij}$ is a ferromagnetic coupling between them and $h_i$ is an external field acting on spin
$i$. The coupling constants $J_{ij}$ equal to a positive constant if spins are connected, and are zero otherwise.

The exact solution for the average spin $\mean{s_i}$ in a single network can be written as \be
\mean{s_i}=\mean{\tanh\left(\beta \sum_j J_{ij} s_j + \beta h_i \right)} \ee where $\beta=1/T$, the temperature
$T$ is measured in units of inverse Boltzmann constant $1/k_B$ and averaging is over the canonical ensemble. If we
now consider an average over all possible realizations of B-A networks, then $\mean{J_{ij}}=J k_i k_j / E$. We use
the mean field approximation, taking $\mean{J_{ij}}$ in place of $J_{ij}$ in our equation. Since $i$ and $j$ are
ordered pairs, the total number of pairs $E$ is twice the number of edges in the network. If we take the external
field $h_i$ equal to zero, the Eq. \ref{startowe} has the following form \be \label{final1} \mean{s_i}=\mean{\tanh
\left(\beta J \sum_j \left( \frac{k_i k_j}{E} s_j \right) \right)} \ee

Now we consider a pair of coupled networks A and B.
The parameters describing both networks can be split into four groups - two describe internal properties of each
network, and two describe network-network interactions. We introduce the following notation: $s_{Ai}$ and $s_{Bi}$
are spins in networks $A$ and $B$, $J_{AA}$, $J_{BB}$ are coupling constants between spins in networks $A$ and $B$
respectively, $J_{AB}=J_{BA}$ are the coupling constants between spins in different networks,
$k_{AAi}$ and $k_{BBi}$ are intra-network node degrees, $k_{ABi}$ and $k_{BAi}$ are inter-network node degrees,
$E_{AA}$ and $E_{BB}$ are twice the total numbers of all intra-network links in $A$ and $B$, $E_{AB}=E_{BA}$ is
the number of links between the networks.\\

Now we extend the Eq. \ref{final1}, introducing the influence of the second network. This way we obtain two
equations for average spins in every network. Since we are interested in critical properties, where average spins
are close to zero, we can approximate the hyperbolic tangent by a linear function and use a standard mean-field
approach. As result we get \ba \mean{s_{Ai}}= \beta J_{AA} k_{AAi} \sum_j \frac{k_{AAj} \mean{s_{Aj}}}{E_{AA}} +
\nonumber \\ \mbox{} + \beta J_{BA} k_{ABi} \sum_j \frac{k_{BAj} \mean{s_{Bj}}}{E_{BA}} \label{raw1} \\
\mean{s_{Bi}}= \beta J_{BB} k_{BBi} \sum_j \frac{k_{BBj} \mean{s_{Bj}}}{E_{BB}} + \nonumber \\ \mbox{} + \beta
J_{AB} k_{BAi} \sum_j \frac{k_{ABj} \mean{s_{Aj}}}{E_{AB}} \label{raw2} \ea To get a relation for the system
critical temperature we need to have a self-consistent equations for order parameter. The case of a single network
required introduction of only single weighted spin $S=\sum_i k_i \mean{s_i} / E$, where the $\mean{s_i}$ is
mean-field average for a given spin $i$. In the case of two connected networks, we need to consider four such
weighted spins $S_{AA}$, $S_{BB}$, $S_{AB}$ and $S_{BA}$. \ba S_{AA}=\sum_i k_{AAi} \mean{s_{Ai}} / E_{AA}
\label{saa} \\ S_{BB}=\sum_i k_{BBi} \mean{s_{Bi}} / E_{BB} \\ S_{AB}=\sum_i k_{ABi} \mean{s_{Ai}} / E_{AB} \\
S_{BA}=\sum_i k_{BAi} \mean{s_{Bi}} / E_{BA} \label{sba} \ea $S_{AA}$ and $S_{BB}$ hold the same meaning as for a
single network, $S_{AB}$ is the mean weighted spin of the network $A$ observed by spins in the network $B$, while
$S_{BA}$ is a mean weighted spin of the network $B$ observed by spins in the network $A$.\\
Eqs.(\ref{weiaa}-\ref{weiba}) were received from Eqs.(\ref{raw1}-\ref{raw2}) by multiplying by appropriate factors (see Eqs.\ref{saa}-\ref{sba}) and summing over $i$. 
These four equations contain only four weighted spins as unknown collective variables and are approximate
mean-field description of the system close to a critical point. It follows we receive
\small \ba S_{AA} = \beta J_{AA} S_{AA} \sum_i \frac{k_{AAi}^2}{E_{AA}} + \beta J_{BA} S_{BA} \sum_i \frac{k_{ABi} k_{AAi}}{E_{AA}} \label{weiaa}\\
S_{BB} = \beta J_{BB} S_{BB} \sum_i \frac{k_{BBi}^2}{E_{BB}} + \beta J_{AB} S_{AB} \sum_i \frac{k_{BAi} k_{BBi}}{E_{BB}} \label{weibb}\\
S_{AB} = \beta J_{AA} S_{AA} \sum_i \frac{k_{AAi} k_{ABi}}{E_{AB}} + \beta J_{BA} S_{BA} \sum_i \frac{k_{ABi}^2}{E_{AB}} \label{weiab}\\
S_{BA} = \beta J_{BB} S_{BB} \sum_i \frac{k_{BAi} k_{BBi}}{E_{BA}} + \beta J_{AB} S_{AB} \sum_i \frac{k_{BAi}^2}{E_{BA}} \label{weiba} \ea \normalsize
If we assume that $k_{ABi}=p_A k_{AAi}$ and $k_{BAi}=p_B k_{BBi}$,
what means that the number of links outside the network is
proportional to the number of links within the network, we can greatly simplify our four equations.
This can be done when one takes into account the way we create inter-network links in our model.\\
Using this assumption, we do not need to consider the cross-network weighted spins $S_{AB}$ and $S_{BA}$ as they are proportional to $S_{AA}$ and $S_{BB}$. Now our first two equations become
\ba S_{A} = \beta J_{AA} S_{A} \sum_i \frac{k_{AAi}^2}{E_{AA}} + \beta J_{BA} S_{B} \sum_i \frac{k_{AAi}^2 p_A}{E_{AA}}\\
S_{B} = \beta J_{BB} S_{B} \sum_i \frac{k_{BBi}^2}{E_{BB}} + \beta J_{AB} S_{A} \sum_i \frac{k_{BBi}^2 p_B}{E_{BB}} \ea
where $S_A\equiv S_{AA}$ and $S_B\equiv S_{BB}$.
The equation array can be written as a single matrix equation.
\be \mathbf{S} = \beta \hat{\Lambda} \mathbf{S} \label{matrix1} \ee

where $\mathbf{S}$ is a vector $\left [\begin{array}{c}S_A\\S_B\end{array}\right ]$ describing the state of the
system and $\hat{\Lambda}$ is a matrix describing effective interaction strengths between spins belonging to the
same or to different networks \footnotesize \be \hat{\Lambda} = \left [ \begin{array}{cc} \Lambda_{AA} & \Lambda_{BA}\\
\Lambda_{AB} & \Lambda_{BB} \\
\end{array} \right ] = \left [ \begin{array}{cc} J_{AA} \frac{\mean{k_{AA}^2}}{\mean{k_{AA}}} & p_B J_{BA} \frac{\mean{k_{BB}^2}}{\mean{k_{BB}}}
\\ p_A J_{AB} \frac{\mean{k_{AA}^2}}{\mean{k_{AA}}} & J_{BB} \frac{\mean{k_{BB}^2}}{\mean{k_{BB}}} \\
\end{array} \right ]. \ee \normalsize

In the case of a single network A, solutions other than $S_A=0$ can exist only if $\beta
J_{AA}\frac{\mean{k_{AA}^2}}{\mean{k_{AA}}}>1$ \cite{bianconi}. In the case of two coupled networks, this
condition corresponds to an eigenvalue of Eq.\ref{matrix1} greater than $1$. The eigenvalues are

\footnotesize \be \label{eigenval} \lambda_{\pm} = \frac{\beta}{2} \left(\Lambda_{AA}+\Lambda_{BB} \pm
\sqrt{\left(\Lambda_{AA}-\Lambda_{BB}\right)^2+4\Lambda_{BA}\Lambda_{AB}}\right) \ee \normalsize

Comparing these eigenvalues with $1$, we get the following critical temperatures
\small \be \label{eq_tc} T_{c\pm}= \frac{\Lambda_{AA}+\Lambda_{BB} \pm \sqrt{\left(\Lambda_{AA}-\Lambda_{BB}\right)^2+4\Lambda_{BA}\Lambda_{AB}}}{2} \ee \normalsize

Since the diagonal elements of $\hat{\Lambda}$, $\Lambda_{AA}$ and $\Lambda_{BB}$ are critical temperatures $T_{cA}$,$T_{cB}$ for separate networks we can write the critical temperatures for the coupled system as
\be T_{c\pm}= \frac{T_{cA}+T_{cB} \pm \sqrt{\left(T_{cA}-T_{cB}\right)^2+4\Lambda_{BA}\Lambda_{AB}}}{2} \ee

To better understand the meaning of these solutions, we introduce the following variables \ba
\mathcal{A}=(T_{cA}+T_{cB})/2 \\ \mathcal{D}=(T_{cA}-T_{cB})/2 \\ \mathcal{C}=\sqrt{\Lambda_{BA}\Lambda_{AB}} \ea
The value $\mathcal{A}$ ("average") describes an average critical temperatures of the networks, $\mathcal{D}$
("difference") is the difference between critical temperatures of both networks, $\mathcal{C}$ ("coupling")
describes a strength of internetwork interactions.\\

Using this notation the critical temperatures can be written shortly as
\be T_{c\pm}=\mathcal{A} \pm \sqrt{\mathcal{D}^2+\mathcal{C}^2} \ee

Let us now consider eigenvectors associated with $\lambda_{\pm}$. They are proportional to the magnetization of
both networks that appears below a given critical temperature and disappears above it. The unnormalized
eigenvectors are \be \mathbf{S}_{\pm}=\left [\begin{array}{c} 1 \\ \frac{-\mathcal{D} \pm
\sqrt{\mathcal{D}^2+\mathcal{C}^2}}{\Lambda_{BA}}\end{array} \right ] \ee The eigenvector $\mathbf{S}_-$ has
opposite signs of its components and corresponds to networks ordered with antiparallel weighted spins, while the
eigenvector $\mathbf{S}_+$ has the same signs of the components and corresponds to networks ordered with parallel
weighted spins.\\ In the limit of vanishing internetwork coupling ($\mathcal{C}=0$) the eigenvalues are simply the
diagonal elements of the matrix $\lambda_A=T_{cA}$, $\lambda_B=T_{cB}$ and the associated normalized eigenvectors
are $\mathbf{S}_A=\left [\begin{array}{c} 1 \\ 0\end{array} \right ]$, $\mathbf{S}_B=\left [\begin{array}{c} 0 \\
1\end{array} \right ]$. This means that in this limit two stable states of the system correspond to the ordering
of just one of the networks, and there is no relation between the order in each networks. It shows our approach
gives correct results in this specific case.\\
Now let us investigate the inequality conditions for existence of solutions $\mathbf{S}_\pm$. From the condition $\lambda_{\pm}>1$ we have
\ba \mathcal{A}-T>\sqrt{\mathcal{D}^2+\mathcal{C}^2} \label{rel1} \\
T-\mathcal{A}<\sqrt{\mathcal{D}^2+\mathcal{C}^2} \label{rel2} \ea

\begin{figure}[ht]
 \vskip 0.5cm
 \centerline{\epsfig{file=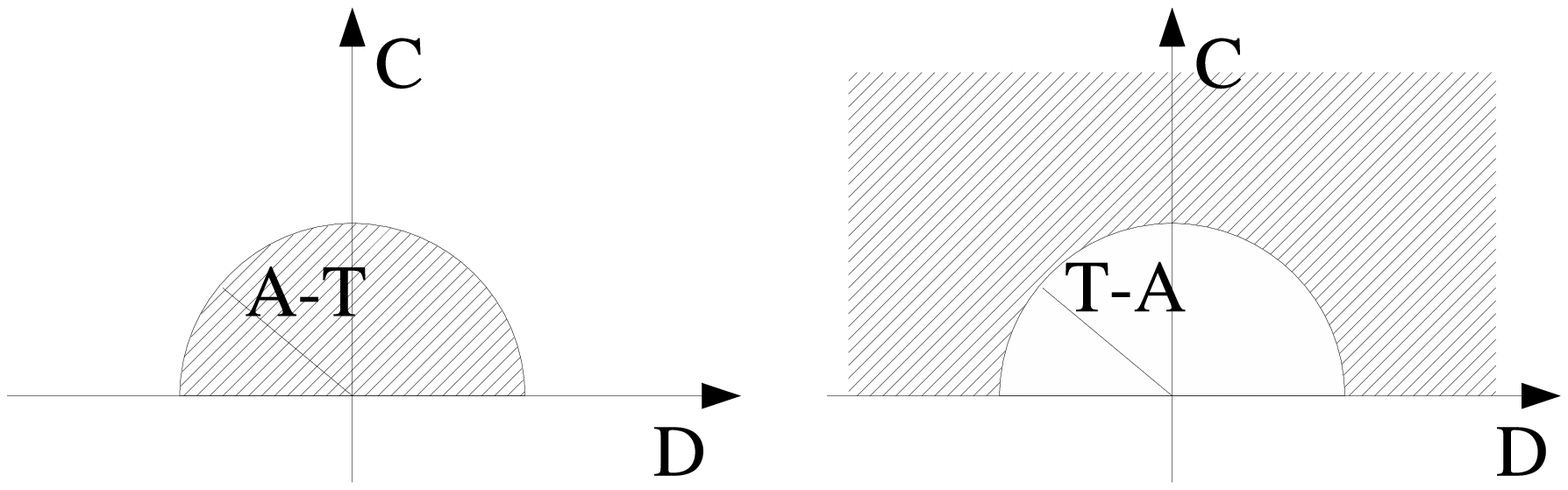,width=.8\columnwidth}}
 \caption{\label{rys_TS} The area of parameters $\mathcal{C}$ and $\mathcal{D}$ where the ordered states $\mathbf{S}_\pm$ can exist. The order appears in the dashed areas. The state $\mathbf{S}_-$ of antiparallel ordering corresponds to the left picture, the state $\mathbf{S}_+$ or parallel ordering corresponds to the right picture. The radius of the circle is the difference $T-\mathcal{A}$. Note that solution $\mathbf{S}_-$ does not exist for $\mathcal{A}<T$ and solution $\mathbf{S}_+$ exist for any $\mathcal{C}$ and $\mathcal{D}$ for $\mathcal{A}>T$. In the case $\mathcal{C}=0$ the networks do not interact at all, and the solutions lose their normal meaning.}
\end{figure}

\begin{figure}[ht]
 \vskip 0.5cm
 \centerline{\epsfig{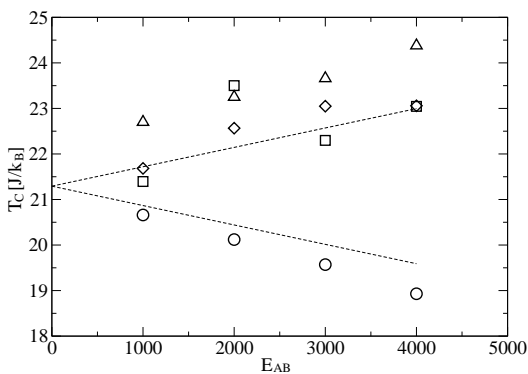}}
 \caption{\label{rys_final} The dependence of temperatures $T_{c\pm}$ on number of links between networks $E_{AB}$. The dashed lines are analytic predictions (Eq.\ref{eq_tc}). The symbols are numeric results. Circles correspond to $T_{c-}$. Squares and diamonds correspond to $T_{c+}$ and are calculated from susceptibility $\chi=(S_h-S_0)/h$. Triangles correspond to $T_{c+}$ and are calculated from susceptibility $\chi \sim \mean{S^2}-\mean{S}^2$.}
\end{figure}

The meaning of these inequalities is presented at Fig.\ref{rys_TS}.\\ If we consider networks of the same size,
the dependence of critical temperatures on inter-network interaction strength $\mathcal{C}$ is linear, and both
critical temperatures are the same for $\mathcal{C} \rightarrow 0$. In such case the system critical temperature
$\mathcal{T}_{c\pm}=\mathcal{A}=T_{cA}=T_{cB}$ (Fig.\ref{rys_final}).\\

The analytic results hold true for any random network, where the probability of a link existing between any two nodes $i$ and $j$ is proportional to the product $k_i k_j$. This is the only assumption about network structure we have used, so any networks where the condition is fulfilled (i.e. random networks) is described by our analysis.\\
Our numeric calculations found in the following section, correspond to the specific case of B-A network and constant coupling $J_{AA}=J_{AB}=J_{BA}=J_{BB}=J$. We can write the critical temperatures as follows
\small \be \label{tc_ba} T_{c\pm}= \frac{T_{cA}+T_{cB}}{2} \pm \sqrt{\left(\frac{T_{cA}-T_{cB}}{2}\right)^2+ p_A p_B T_{cA} T_{cB}} \ee \normalsize
where $T_{cA}=J (m_A/2) \ln{N_A}$ and $T_{cB}= J (m_B/2) \ln{N_B}$. The values of $p_A$ and $p_B$ are not independent, and are connected with the number of links between networks $p_A E_{AA} = p_B E_{BB} = E_{AB}$.\\

\section{Numeric results}

Our analytic calculations show the existence of two different ordered states and estimate values of two critical
temperatures where these states disappear. Below we investigate numerically a case of two coupled B-A networks
with the same number of nodes $N_A=N_B=5000$ and links $E_{AA}=E_{BB}=40000$ ($m_A=m_B=5$,
$\mean{k_{AA}}=\mean{k_{BB}}=10$).\\

We run Ising dynamics on these networks, setting the following initial condition: all spins in both networks have
the same value $s_{Ai}=s_{Bi}=+1$. We allow the system to relax for $\tau=20$ time steps, perform averaging for
$\tau=20$ time steps, then increase the temperature and start from the same initial condition as before. This way,
results for different temperatures are not correlated. We find the weighted spin $S=S_A+S_B$ for each
temperature $T$ and average it over $100$ network realizations. Simulations are performed for different numbers of
inter-network links $E_{AB}=1000,2000,3000$ and $4000$ that were attached preferentially to fulfill the
assumptions of our model (see the discussion at the end of Sect.\ref{dysk}).\\

\begin{figure}[ht]
 \vskip 0.5cm
 \centerline{\epsfig{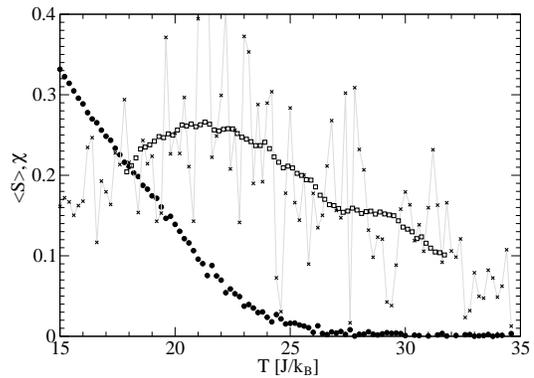}}
 \caption{\label{rys_same}The dependence of the average weighted spin $S=S_A+S_B$ and its susceptibility for small external field $h=0.05 J$ in the case of two B-A networks with $E_{AB}=1000$ connections between them. An initial condition for each temperature is a completely ordered system. The full symbols depict $S$, the X symbols correspond to susceptibility $\chi=\left(S_{0.05}-S_0\right)/h$ (the lines are just to guide eye), the empty symbols are 30-point running average of the susceptibility. The parabolic fit was used to find the susceptibility maximum.}
\end{figure}

For low temperatures $T$ the system is ordered. As the temperature increases the average weighted spin $S$ decreases.
When the temperatures increase over $T_{c+}$, the ferromagnetic ordered state of both networks disappears, i.e.
both networks become paramagnetic (Fig.\ref{rys_same}).\\

Finding the exact value of critical temperature $T_{c+}$ from numerical simulations is not straightforward. If one
observes the dependence of the weighted magnetization on rising temperature and tries to fit the magnetization
decay to a linear or to an exponential function the results strongly depend on relaxation time $\tau$. To overcome
this problem we observed the temperature dependence of the system {\em susceptibility} $\chi$. In fact, by
comparison to standard models of magnetic systems, one can expect that the initial susceptibility diverges at
$T=T_{c+}$. In our finite system we are looking simply for the maximum of $\chi$. To estimate $\chi$ we are using
two methods. First we compare average weighted spin $S$ for a small external field $h=0.05 J$ and the value of
$S_0$ with no external field. It follows $\chi=\left(S_h-S_0\right)/h$. Because such results are strongly
fluctuating as a function of system history and temperature (Fig.\ref{rys_same}), we calculate running average
over $30$ temperature points and find the maximum of $\chi$ by fitting a parabolic curve. The top of the parabola
corresponds to the position of the critical temperature $T_{c+}$. We found that these values are independent on
the relaxation time $\tau$ used in our numerical experiment. The second method of finding the critical temperature
$T_{c+}$ is observation of the time average $\mean{S^2}-\mean{S}^2$, where we average over one relaxation period
$\tau$. The magnitude of the fluctuations is proportional to the suceptibility $\chi \sim \mean{S^2}-\mean{S}^2$
according to the fluctuation-dissipation theorem \cite{flucdis}. Similarly to previous method, we calculate running average over
$10$ points and find the maximum. The values are shifted by a constant value comparing to analytic results and do
not fluctuate as much as those obtained from the first method (see Fig.\ref{rys_final}).\\

\begin{figure}[ht]
 \vskip 0.5cm
 \centerline{\epsfig{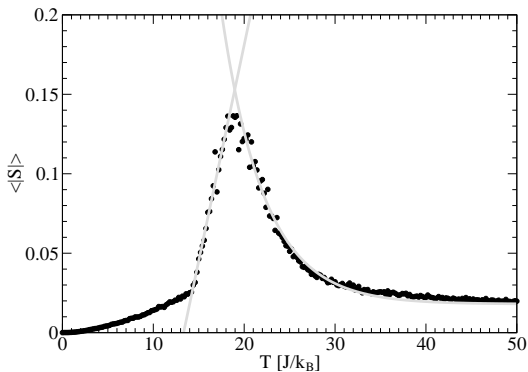}}
 \caption{\label{rys_bump}The temperature dependence of the average weighted spin value $\abs{S}=\abs{S_A+S_B}$ in the case of two B-A networks with $E_{AB}=4000$ connections between them. Initial conditions for each temperature are two fully ordered networks with opposite spins. The symbols correspond to numerical data and grey lines are approximations of magnetization behavior below and above critical temperature.}
\end{figure}
\begin{figure}[ht]
 \vskip 1cm
 \centerline{\epsfig{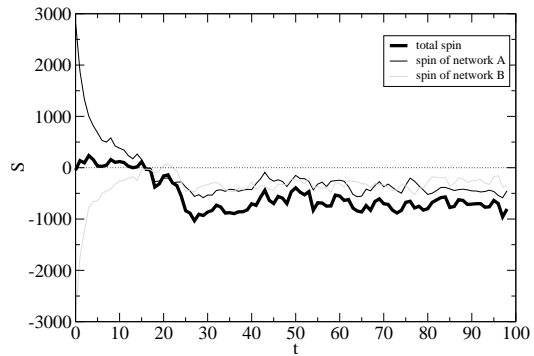}}
 \caption{\label{rys_example}An example of time evolution of magnetization for two networks of size $N_A=N_B=5000$ and $\mean{k}=4$ with $E_{AB}=500$ links between them in temperature $T=30$ (scaled in $k_B/J$). At this temperature, the antiparallel ordered state is not stable, and the initial condition of networks ordered in such fashion evolves towards a stable parallel ordered state.}
\end{figure}

Now we consider the same networks with the following initial condition for each temperature: spins in both
networks are ordered antiparallel $s_{Ai}=+1, s_{Bi}=-1$. The relaxation algorithm and the measurement procedure
is the same as for parallel spin case, however now we consider absolute values of weighted spin of the whole
system $\abs{S}=\abs{S_A+S_B}$. For low temperatures $T$ both networks remain ordered in opposite directions
(Fig.\ref{rys_bump}) and as result an average weighted spin value $\abs{S}=\abs{S_A+S_B}$ fluctuates around $0$.
When the temperature increases over $T_{c-}$, the state of antiparallely ordered networks disappears and networks
start to order in a parallel fashion. An example of such a scenario is presented at Fig.\ref{rys_example}. When
temperature further increases and $T>T_{c+}$ the system becomes paramagnetic.\\

We found numerically the critical temperature $T_{c-}$ from the intersection of extrapolations of rising and
declining part of the curve $S(T)$ (linear fit was used for the rising part and exponential fit for the declining
part) (Fig.\ref{rys_bump}).
This point slightly depends on relaxation times $\tau$. For large $\tau$ the effect of finite system size can be
easily observed and the system jumps from an antiparallel state to a parallel one that has a lower energy.\\

The results for dependence of $T_{c-}$ and $T_{c+}$ on the number of links between networks $E_{AB}$ agree with
the analytic calculations (Fig.\ref{rys_final}).

\section{Conclusions}
In the system of two B-A network, the Ising model possesses two low-temperature stable states --- both networks
ordered parallel or antiparallel. It follows there are two critical temperatures corresponding to the
disappearance of these two stable states --- $T_{c-}$ and $T_{c+}$. They are placed symmetrically around the
average of critical temperatures of separate networks. The difference between them depends on density of
inter-network links and the difference between critical temperatures of separate networks. The analytic
calculations agree with performed numeric simulations.

\begin{acknowledgments}
This work was partially supported by a EU Grant {\it Measuring and Modelling Complex Networks Across Domains}
(MMCOMNET) and by State Committee for Scientific Research in Poland (Grant No.1P03B04727).
\end{acknowledgments}

\end{document}